# A Design Methodology for Space-Time Adapter

CHAVET Cyrille[1], COUSSY Philippe[2], URARD Pascal[1], MARTIN Eric[2]
[1]STMicroelectronics, Crolles, FRANCE. {firstname.lastname@st.com}
[2]LESTER Lab, UBS University, CNRS FRE 2734. {firstname.lastname@univ-ubs.fr}

**ABSTRACT**
This paper presents a solution to efficiently explore the design space of communication adapters. In most digital signal processing (DSP) applications, the overall architecture of the system is significantly affected by communication architecture, so the designers need specifically optimized adapters. By explicitly modeling these communications within an effective graph-theoretic model and analysis framework, we automatically generate an optimized architecture, named Space-Time AdapteR (STAR). Our design flow inputs a C description of Input/Output data scheduling, and user requirements (throughput, latency, parallelism…), and formalizes communication constraints through a Resource Constraints Graph (RCG). The RCG properties enable an efficient architecture space exploration in order to synthesize a STAR component. The proposed approach has been tested to design an industrial data mixing block example: an Ultra-Wideband interleaver.

**Categories and Subject Descriptors**
B6.3 [**Design aids**]: Automatic synthesis, Optimization.

**General Terms**
Algorithms, Design & Performance.

**Keywords**
Communication and interface synthesis, RTL design, Digital Signal Processing and Multimedia Applications.

## 1. INTRODUCTION

The ever growing complexity of applications and the shrinking time-to-market lead the designers to look for advanced design methodologies. Indeed, to design such complex architecture within a short design time, it is necessary to raise the abstraction level of design description to system level, to explore the design space and finally to automatically generate the hardware register transfer level (RTL) architecture. Nowadays, a widespread solution to handle design complexity is to reuse pre-design heterogeneous IP cores. Unfortunately, the main problem arises from their integration.

In the multi-processor SoC (MPSoC) context (IP cores can be processor, memory, bus…) the problems come from the interfaces and protocols of the components. To tackle interfacing and functional problems when designing MPSoC architectures, system integrators can use standard interfaces such as Virtual Component Interface proposed by VSIA [16] and Open Core Protocol proposed by the OCP International Partnership [17]. However, in addition to the protocol aspects, SoC designers also have to synchronize components and to buffer data in order to ensure system behavior and to meet timing constraints. In [7] authors propose to automatically generate simulation wrappers for MPSoC architectures.

However, in the field of Digital Signal Processing (DSP) applications (e.g. [13]), a multi-processor SoC (MPSoC) architecture may not be a well-suited solution because of design complexity. Optimized hardware accelerators (e.g. filters) - composed of a set of computing blocks communicating through point-to-point links- are still needed. From this point of view, the designers have to tackle problems such like throughput adaptation, data re-ordering (e.g. row-column), Input/Output parallelism adaptation. Based on communication templates, [9] presents a generic interface unit architecture for communication synthesis in a *platform-based* design approach. In [1] a multiplexer/demultiplexer and FIFO-based interface architecture is used. In [6], the authors propose a systematic way of interfacing data-flow hardware accelerators (IP core) for their integration in a system on chip. Their interface architecture is based on FIFO (queue) storage elements and a Direct Memory Access module (DMA). They assume that the IP are data synchronized (i.e. at each clock cycle a data is presented and read). However, these previous approaches assumed that the sequence of produced data is the same as the sequence of consumed data (no re-ordering). Moreover, FIFO sizes are computed by a "set and simulate" approach.

Obviously, interfacing DSP's blocks greatly impacts the quality of the system (throughput, area, power consumption…), that's why efficient communication adapter design is still one of the most important points in complex system design. In fact, using Input/Output (I/O) wrappers can introduce unnecessary memorizing elements. Such wrappers may be needed in order to solve data reordering problems that can arise from the IP core integration. In [12] the authors aim at determining at compile time whether a FIFO is sufficient for every producer/consumer pair of a Kahn Process Network. When the sequence of produced data is different from the sequence of consumed data, extra storage and control on the consumer side is proposed [15]. This extra module includes a CAM (Content Addressable Memory) where data are addressed using a hash table. This solution enables the implementation of non-deterministic communications, but there is no optimization of the adapter overhead since overlapping of input and output data is not possible. In [2], a formal technique for hardware interface design is proposed. A generic interface model targeted by the communication synthesis is used. The low-level timing constraints can include strict timing specifications or data transfer schedule. The interface synthesis is carried out by an allocation procedure of data storage components (FIFO, LIFO and register). However, the size of storage elements is not computed or even taken into account during the design process. The proposed methodology is based on NP-complete maximum clique algorithm. In [14] the authors develop a system-level IP reuse methodology where designs are described in three layers. Data transfer and data storage optimizations are done by reorganizing loop indexing and loop nesting. Unfortunately, the authors do not present the technique they use to produce the RTL component architecture from the algorithm specification. In [4], the authors develop a set of techniques dedicated to the design of DSP algorithm. High-level synthesis of the processing unit is carried out under I/O timing and architectural constraints. The approach leads to an optimized data-path synthesis but still requires the communication unit design.

In [11] authors proposed approaches that use Matlab/Simulink for the system specification and that produce a VHDL RTL architecture of the system. Based on hardware macro generators that use the "generic"/"generate" mechanisms, the synthesis process can be summarized as a block instantiation and block interconnection thanks to memory blocks. However, minimizing such buffer memory size in automatic code generation from the high-level system specification is still one of the key technologies [8]. That's why in [8] the authors propose a methodology for the reduction of on-chip memory size. Our goal is to tackle the same problem, but our methodology analyses the communication at a finer grain level. This fine grain communication analysis enables deep exploration of optimization solutions and helps us to generate a close to the best memorization architecture.

In this paper, we present an automatically generated optimized Space-Time AdapteR (STAR). Our design flow inputs timing diagrams (constraints file) or a C description of I/O data scheduling (e.g. an interleaving formula), and user requirements (throughput, latency…), and formalizes communication constraints through a formal Resource Constraints Graph (RCG). The RCG properties enable an efficient architecture space exploration in order to synthesize a STAR component. The contribution of our work can be seen as a solution for the automatic generation of a static network on chip. Indeed, our Space-Time AdapteR (STAR) architecture can be used to interconnect a set of IP cores.

The paper is organized as follows: the second section is dedicated to the problem formulation. In the third section we present our design flow, while the associated formal models and methodology are detailed in section four. Finally, the last section presents experimental results.

## 2. MOTIVATION

Let us consider a simple architecture example composed of two components exchanging a set of data S = {a, b, c, d, e, f}. S is produced by a block #1 and is consumed by a block #2 through a single point-to-point link.

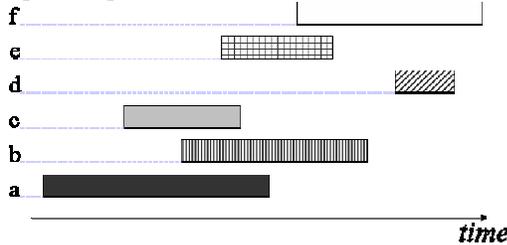

Figure 1: Data lifetime.

The write access sequence into the communication link is $S_w$ = (a,c,b,e,f,d) i.e. $t^w_a < t^w_c < t^w_b < t^w_e < t^w_f < t^w_d$, while the read access sequence from the link is different $S_r$ =(c,a,e,b,d,f) i.e. $t^r_c < t^r_a < t^r_e < t^r_b < t^r_d < t^r_f$ (see Figure 1). This difference between the two I/O sequences can either come from the integration of two IP cores that were not specifically designed to work together, either can be explicitly described (e.g. in interleavers [5][18]). As those blocks do not produce and consume data in the same order nor with the same throughput (nor sometime the same parallelism), they can not be directly plugged together. The designer needs to introduce a space-time adapter between them to ensure correct functional results. A classical solution consists in using a memory to buffer all concerned data: this is what we call coarse grain approach. But in fact, this over sized buffer may be reduced thanks to a finer grain communication constraints analysis [4]. The proposed adapter can be designed either by using a set of registers or specific memory elements, such as FIFO (queue) or LIFO (stack). The problem the designer faces consists in finding the best architecture for this adapter: he has to find the best storage element binding.

For example, the lifetimes of data *a* and *b* respect a First-In First-Out semantic, so they can be assigned to the same hardware FIFO. This timing relation is also true for the data *c* and *b*. However, data *a* and *c* respect a Last-In First-Out semantic, so a single hardware FIFO cannot be used to store the data *a*, *b* and *c* The question for the designer is: how can we bind data *a*, *b* and *c* to different storage elements, in order to generate the best final architecture? This highlights the fact that the local problem of *a*, *b* and *c* binding will influence the resulting global architecture. A methodology is thus needed to bind data *a*, *b* and *c* to different storage elements, in order to generate an optimized architecture.

## 3. PROPOSED SOLUTION

The architecture of a STAR component is composed of a datapath and the associated control state machine FSM (see Figure 2). The data path can be composed of FIFO, LIFO or register. Spatial adaptation (a data read on one input port can be send to any/several output ports) is performed by an interconnection logic dealing with data dispatching from input port to storage elements, and from storage elements to output ports. We can see on Figure 2 that there is one STAR architecture for each input port.

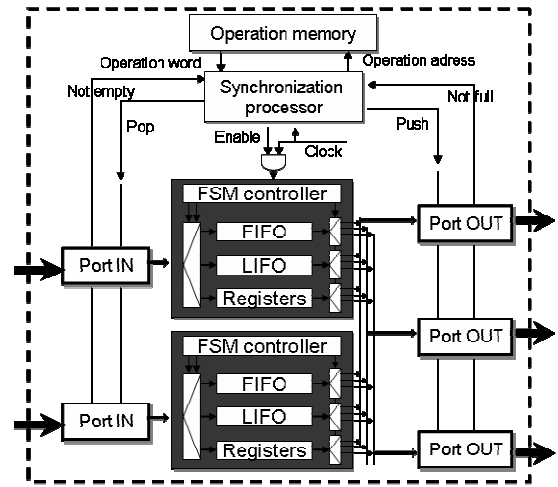

Figure 2: Typical STAR architecture.

The timing adaptation (data-rates, different input/output data scheduling) is realized by the storage elements. STAR can have a GALS (Globally Asynchronous Locally Synchronous) / LIS (Latency Insensitive System) interface as described in [3].

The design flow is presented in Figure 3 and is currently based on three tools: *StarTor* for the STAR design constraint specification, *StarGene* for the STAR component synthesis and *StarBench* for the STAR functional validation. The methodology generates a register transfer level (RTL) architecture starting from a functional model and a set of user requirements (timing and communication-architecture constraints). The architecture synthesis is performed by using a library of pre-designed and characterized storage elements (FIFO, LIFO and Registers).

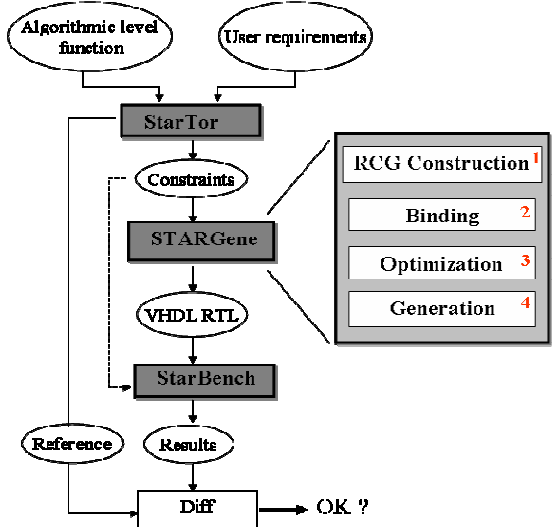

Figure 3: STAR design flow and associated tools.

*StarTor* inputs a C level algorithmic description which specifies the interleaving scheme, and a file containing user requirements (latency, throughput, communication interface, I/O parallelism...). StarTor first extracts I/O data communication order by generating a trace from the execution of the C functional description. Next, based on designer's requirements, it generates a constraints file. This file contains the number and type of ports, type and amount of data, relationships between data and ports (i.e. mapping) and finally read and write access dates for all data. Then, in order to generate a STAR component, our design tool *STARGene* is based on a four-step flow: (1) Resource Compatibility Graph construction, (2) Storage resource binding, (3) Architecture optimization and (4) VHDL RTL generation (see Figure 3). During the first step of the STAR component Generation, a Resource Constraints Graph RCG is generated from the communication constraints. The analysis of this formal model allows both data binding to storage elements (queue, stack or register), and the sizing of each storage element. This first architecture is next optimized by merging storage elements that have non-overlapping usage timing frames. Finally, an RTL level design is generated. The last tool, *StarBench*, generates a test bench based on constraints in order to validate the design by comparing simulation results.

Typically, a STAR could have to deal with different execution modes (configuration), switching from one to another at run-time. In this paper, we present a formal methodology to synthesize a STAR architecture for a given configuration. The generalization of the methodology generating multi-mode architecture (graph merging, multi data path synthesis, multi FSM generation…) will be presented in a future publication.

## 4. STAR DESIGN FLOW

### 4.1. Resource Compatibility Graph Construction

The first step consists in generating a Resource Compatibility Graph, from the design constraints file. This RCG specifies through formal modeling the timing relationship between data that have to be handled by the STAR architecture. The vertex set $V=\{v_0, ..., v_n\}$ represents data, the edge set $E=\{(v_i, v_j)\}$ represents the compatibility between the vertices. A tag $t_{ij} \in T$ is associated with each edge $(v_i,v_j)$. This tag represents the *compatibility type* between the two data ($i$ and $j$), $T= \{Register\ R,\ FIFO\ F,\ LIFO\ L\}$, *e.g.* Figure 4.

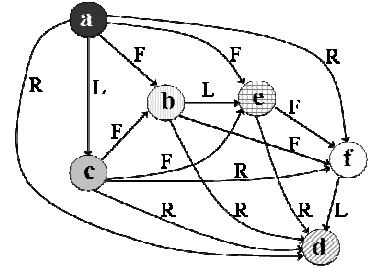

Figure 4: Graph example (from Figure 1 constraints).

In order to assign compatibility tags to edges, we need to identify the timing relationship that exists between two data. For this purpose we defined a set of rules based on functional properties of each storage element (FIFO, LIFO, Register).

The *lifetime* of data $a$ in a **STAR** is defined by $\Pi(a) = [\tau_{min}(a),\ \tau_{max}(a)]$ where $\tau_{min}(a)$ and $\tau_{max}(a)$ are respectively the date of the write access of $a$ into the component and the last date of the read access to $a$. $\tau_{first_a}$ is the first read access to $a$, $\tau_{Ri_a}$ is the $i$-th read access to $a$ with $first \leq i \leq max$.

***Rule 1***: *Register compatibility*
If $(\tau_{min_b} \geq \tau_{max_a})$ then we create a "Register" tagged edge.
Here, data lifetime intervals are said to be "*un-overlapping*". In other words, those two data can be stored in the same storage element.

***Rule 2***: *FIFO compatibility*
If $[(\tau_{min_b} > \tau_{min_a})$ and $(\tau_{first_b} > \tau_{max_a})$ and $(\tau_{min_b} < \tau_{max_a})]$ then we create a "FIFO" tagged edge.
In this case, data lifetime intervals are said to be "*partially overlapping*" and data $a$ and $b$ can be stored in the same FIFO structure; Note that the last relation $(\tau_{min_b} < \tau_{max_a})$ enables a formal distinction with *Register* compatibility. The FIFO structure size is not always equal to the maximum number of data stored in it. This point will be detailed in the next section.

***Rule 3***: *LIFO compatibility*
If $[[(\tau_{min_b} > \tau_{min_a})$ and $(\tau_{first_a} > \tau_{max_b})]$ or $[(\tau_{Ri_a} < \tau_{min_b} < \tau_{max_b} < \tau_{Ri+1_a})]]$ then we create a "LIFO" tagged edge.
In this case, data lifetime intervals are said to be "*including-overlapping*". In the rest of this paper, we will only consider the first part of this rule, i.e. $[(\tau_{min_b} > \tau_{min_a})$ and $(\tau_{first_a} > \tau_{max_b})]$. In this case, LIFO structure size equals the maximum number of data stored in it. Future works will integrate the complete rule in our tool.

***Rule 4:*** Otherwise, *No edge - No compatibility*
In this case, we say the data are *incompatible*: two different elements have to be used to store data $a$ and $b$.

An analysis of I/O timing relations, we generate a RCG. The graph construction supposes edge creation between data, respecting a chronological order ($\tau_{min}$). If $n$ is the number of data to be handled, the graph may contain: $n(n-1)/2$ edges, $O(n^2)$.

### 4.2. Storage element binding

The second step consists in binding storage elements to data by using the timing relations modeled by the RCG. The aim is to identify and to bind as many FIFO or LIFO structures as possible on the RCG.

In [2], by searching and isolating compatibility cliques in an undirected graph, the authors identify the different storage structures (FIFO or LIFO). This approach has four main drawbacks: (1) identifying a maximum clique in an undirected graph is a NP-complete problem (*resource identification step*), (2) when such a clique is found, analysis have to be performed to define the clique type (FIFO or LIFO) and to check if the I/O constraints are respected (*resource identification step*), (3) the proposed flow does not allow sizing of identified storage elements (*resource sizing step*) and (4) the authors do not propose any exploration algorithm (*resource binding step*).

**Resource identification:**

In our approach, the type of structures needed to handle two data is modeled by a tagged edge. Then, the storage element identification (FIFO, LIFO or Register) is made easier (polynomial algorithm) by using the notion of *path*. Traveling a path of a given type (F or L) with RCG modeling is equivalent by construction, to the compatibility clique searching described in [2].

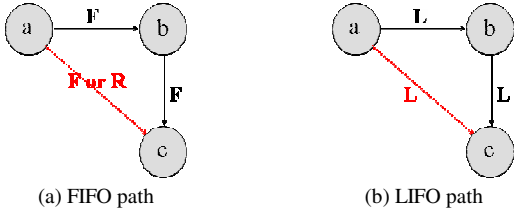

(a) FIFO path  (b) LIFO path
Figure 5. Compatibility cliques identification.

Let $a$, $b$, $c$ be three chronologically ordered FIFO compatible data ($\tau_{min_a} < \tau_{min_b} < \tau_{min_c}$),

*Theorem 1*
**If $a$ is FIFO compatible with $b$ and $b$ is FIFO compatible with $c$, then $a$ is transitively FIFO (or Register) compatible with $c$.**

Due to space limitation, the formal proof of this theorem will not be given here, but it can be easily proven using the definition of FIFO compatibility, and thanks to the transitivity of the inequality relation. However, the distinction between FIFO and Register compatibility ($\tau_{first\ b} > \tau_{max_a}$) in the definition of FIFO compatibility cannot be transformed by transitivity. Since it is used to distinguish F and R compatibility, we do not have enough information to make this distinction in the resulting edge. So the compatibility between $a$ and $c$ can be FIFO or Register (Figure 5.a).

*Lemma 1:* A FIFO compatible data path $P_F$ is, by construction, a compatibility clique corresponding to a set of data that can be stored in a single FIFO.

This can be proven by a recursive application of *Theorem 1* on $P_F$.

*Theorem 2*
**If $a$ is LIFO compatible with $b$ and $b$ is LIFO compatible with $c$, then $a$ is transitively LIFO compatible with $c$.**

Due to space limitation, the formal proof of this theorem will not be given here, but it can be easily proven using the definition of LIFO compatibility, and using the transitivity of the inequality relation. Then data $a$ and $c$ are said to be LIFO compatible by definition (Figure 5.b).

*Lemma 2:* A LIFO compatible data path $P_L$ is, by construction, a compatibility clique corresponding to a set of data that can be stored in a single LIFO.
This can be proven by a recursive application of *Theorem 2* on $P_L$.

**Resource sizing:** The size of a LIFO structure equals the maximum number of data stored by a LIFO compatible data path. So, we have to identify the longest LIFO compatibility path $P_L$ and then the number of vertices in $P_L$ equals the maximum number of data that can be stored in this LIFO (see Figure 6).

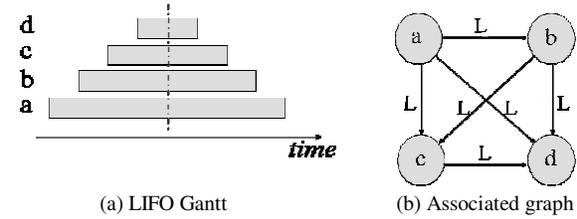

(a) LIFO Gantt  (b) Associated graph
Figure 6: LIFO compatibility cliques.

However, data from a FIFO compatible path are not always FIFO compatible with each other (e.g. Figure 7.a). So the size of a FIFO structure is not always equal to the number of data in the path: the size of the FIFO is the maximum number of data (of the considered path) stored at the same time in the structure. In fact, the aim is to count the maximum number of overlapped data (respecting I/O constraints) in the selected path $P$.

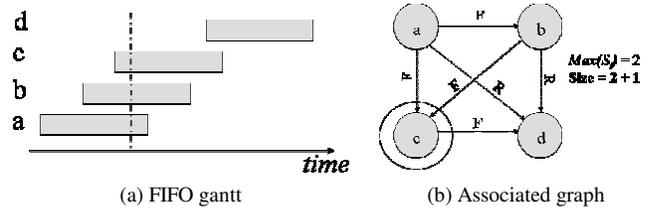

(a) FIFO gantt  (b) Associated graph
Figure 7: FIFO compatibility cliques.

*Theorem 3*
Let $P$ be the longest FIFO compatibility path (edges tagged with F),
Let $i$ be a vertex of the graph, remaining in $P$,
Let $S_i$ = number of incoming FIFO tagged edges, whose origin vertex is in $P$,
Then, **Size = 1 + max ({ $S_i$ | for all vertices $i$ in $P$})**.

**Resource binding:** We use a greedy algorithm based on user plotted metrics (minimal amount of data to use a FIFO or a LIFO, average use factor, FIFO/LIFO usage priority factor, complexity of routing architecture…), to bind as many FIFO or LIFO structures as possible on the RCG. A two-steps flow is used: (1) identification of the best structure, (2) merging all the concerned data in a hierarchical node. Then, each node represents a storage element, as shown on Figure 8.a (e.g. data *a*, *b* and *f* are merged in a 3-stages FIFO). We say *hierarchical node* because merging a set of data in a given node, supposes adding information that will be useful during the optimization step: the lifetime of this structure (i.e. the time interval during which this structure will be used. e.g. Figure 8.b).

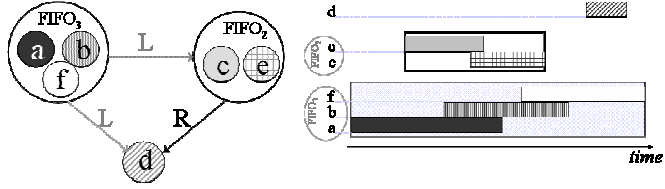

(a) Resulting hierarchical graph      (b) Resulting constraints
Figure 8: A possible binding for Figure 4 graph.

Let P = $\{v_0, ..., v_n\}$ be a compatible data path,
- If P is a FIFO compatible path, the structure lifetime will be $[\tau_{min_{v_0}}, \tau_{max_{v_n}}]$,
- If P is a LIFO compatible path, the structure lifetime will be $[\tau_{min_{v_0}}, \tau_{max_{v_0}}]$.

The selection of the nodes to be merged in a hierarchical one influences the resulting architecture, since these nodes will not be used to build another structure. When such a structure (FIFO or LIFO) has been identified, i.e. when the corresponding hierarchical node has been created, the binding step exploration is performed on the rest of the graph. When no more FIFO or LIFO structures can be identified on the graph, the next step is architecture optimization.

### 4.3. Architecture Optimization.

The goal of this task is to maximize storage resource usage and buffer sharing, in order to optimize the resulting architecture by minimizing the number of storage elements and the number of structures to be controlled. The goal is to merge, if possible, the previously bound structures.

To tackle this problem, we build a new hierarchical RCG with these hierarchical nodes, and their lifetimes. In order to avoid any conflict, the exploration algorithm of the optimization step will only search for Register compatibility path (buffer with disjoints lifetimes), between same type vertices. When two or more structures of the same type are Register compatible all together, they can be merged.

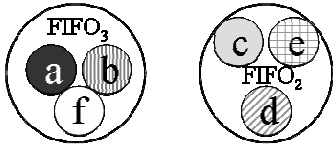

Figure 9: Optimization of Figure 8 graph.

Let P = $\{v_0 ... v_n\}$ be a Register compatible data path,
- The lifetime of the resulting hierarchical merged structure will be $[\tau_{min_{v_0}}, \tau_{max_{v_n}}] \cup ... \cup [\tau_{min_{v_n}}, \tau_{max_{v_n}}]$.

The algorithm is very similar to the one used during binding step. When there is no more merging solution, the resulting graph is used to generate the RTL VHDL architecture. Figure 9 is a possible solution for the constraint set presented in Figure 1.

Here, the resulting architecture consist in a 3-stages FIFO that handles 3 data, and a 2-stages FIFO that handles 3 data: one memory place has been saved.

## 5. EXPERIMENTS

In this section we show the results of using our design flow to generate (1) a STAR architecture based on FIFO storage elements compared to a STAR architecture based on a sea of registers, (2) an Ultra Wide Band interleaver [18] example. We use DCUltra Synopsys for logic synthesis from the generated RTL STAR architecture. All the areas have been masked and we also use arbitrary units (To protect STMicroelectronics technologies).

### 5.1 In-order transaction study

In order to highlight the interest of FIFO/LIFO structures in STAR components, we first generate a naïve architecture based on a single FIFO; storing various numbers (from 32 up to 288) of 8 bit data (see Table 1). Next we compare it to the corresponding architecture using a "sea" of registers generated by our tools (using the "register only" option). The corresponding constraints file specifies that the data are read (written) one by one through (on) one input (output) port, and no data can be read before all data are stored (no overlapping between inputs and outputs). Thus, the resulting architecture stores all data, preventing any optimization.

Table 1 show that the control area for the FIFO-based architecture is smaller than the control area for a register-based STAR when the total amount of data increases. This result comes from the number of storage elements to be controlled: on the one hand, one FIFO and on the other, a sea of registers. The difference between data path areas arises from the greater integration density of the RAM blocks that are used to design FIFO/LIFO structures.

Table 1. Area results for reference test case (a.u.²)

| # data | FIFO-based | | | Register-based | | |
|---|---|---|---|---|---|---|
| | Data path | Control | Total | Data path | Control | Total |
| 32 | *5888* | *1511* | **7399** | *7040* | *3258* | **10298** |
| 64 | *7860* | *1522* | **9382** | *14080* | *4959* | **19039** |
| 128 | *12276* | *1561* | **13837** | *28160* | *8539* | **36699** |
| 256 | *18672* | *1588* | **20260** | *56320* | *16061* | **72381** |
| 272 | *20052* | *1675* | **21727** | *59840* | *17597* | **77437** |

### 5.2 Ultra-Wide Band interleaver

This component has to be able to switch between different modes (300, 600 or 1200 data length), respecting latency constraints. By nature, interleavers are nearly worst case test-benches for our design flow, since they offer few storage elements to be saved. In a simplistic way, the more the data are interleaved; the better the functional results are for telecommunication applications. However, these data-mixing schemes are well-known and very pedagogical mathematical examples and we can explore how metrics (I/O parallelism, enable/disable FIFO/LIFO, average usage factor…) can influence the final architecture.

In Table 2, the number in column *saved* is the number of register saved, and the number in *Ctrl* column is the number structure to be managed. These results has been obtained with a parallelism of 6 data input and 10 data output. Additional constraints used during synthesis are F/L minimum length (e.g. 7 or 15) and filling (%).The reference design from STMicroelectronics has been generated using a commercial HLS tool. We also use our tools to generate the corresponding architecture based on a sea of register (No F/L) for each mode. In the reference architecture there is no memory saving (1200 registers in the worst case, 2400 when pipelined) but the three modes are integrated in a single architecture.

Using our flow, we can save registers and decrease latency in any case. Moreover the number of structure to be controlled is smaller when we use our model. Drawback of this result is that the reduction of storage elements can increase the complexity of data multiplexing (depending on the interleaving rule). However our approach also enables to enhance the throughput by optimizing the latency to input and next output data. So, depending on the selected mode the throughput of our architecture can vary from 412 to 438 Mb/s (related to Table 2 designs) compared to 375Mb/s as a theoretic throughput from the reference (Table 2).

**Table 2. Compared results for a given I/O parallelism**

| Mode | Reference | | F/L (Min 7 / 95%) | | F/L (Min 15 / 90%) | | No F/L | |
|------|-----------|------|-------|------|-------|------|-------|------|
|      | Saved     | Ctrl | Saved | Ctrl | Saved | Ctrl | Saved | Ctrl |
| 300  | n/a       | 300  | 56    | 77   | 60    | 240  | 60    | 240  |
| 600  | n/a       | 600  | 83    | 101  | 130   | 470  | 130   | 470  |
| 1200 | n/a       | 1200 | 96    | 117  | 120   | 609  | 168   | 1032 |

Currently, we generate the different modes separately, while the reference design integrates the three modes in a single 2400 memory points design. But when we concatenate our three designs (one for each mode) in a single architecture, the total area is about 14% smaller than the reference design. Future works will enable the generation of optimized multi-modes architectures to further reduce the area.

## 6. CONCLUSION

In this paper, we proposed a design space exploration methodology for Space-Time AdapteR STAR components. This approach relies on the formal modeling of communication constraints based on a Resource Compatibility Graph RCG describing timing relations between data. The binding and optimization steps that assign data to storage elements according to the timing relations have been presented. Experimental results in the telecom domain have demonstrated the interest of this methodology. Formal modeling allows RTL architectures to be synthesized from a single C functional specification and under various I/O timing constraints. We also show that it is easy to explore different solution by applying different constraints during synthesis. This allows enhancements based on refinements.

Future works will focus on the formal transformation of the RCG in order to generate multi-configuration and pipelined architectures. Moreover, we also investigate the use of a STAR architecture in high-level synthesis flow: in our flow we use scheduling information –available from a high level synthesis tool- about data accesses and the cycles that they occur in. Then the STAR can be use to implement computation memory [10].